\documentclass[journal]{IEEEtran}
\usepackage{cite}
\usepackage{amsmath}
\usepackage{graphicx}
\usepackage{amssymb}
\usepackage{mathdots}
\usepackage{multirow}
\usepackage{arydshln}
\usepackage[para,online]{threeparttable}

\usepackage{enumitem}
\usepackage{stfloats}

\begin{document}
\title{Application of Distributed Arithmetic to Adaptive Filtering Algorithms: Trends, Challenges and Future}
\author{Mohd.~Tasleem~Khan 
}

\maketitle
\begin{abstract}
The utilization of distributed arithmetic (DA) in AF algorithms has gained significant attention in recent years due to its potential to enhance computational efficiency and reduce resource requirements. This paper presents an exploration of the application of DA to adaptive filtering (AF) algorithms, analyzing trends, discussing challenges, and outlining future prospects. It begins by providing an overview of both DA and AF algorithms, highlighting their individual merits and established applications. Subsequently, the integration of DA into AF algorithms is explored, showcasing its ability to optimize multiply-accumulate operations and mitigate the computational burden associated with AF algorithms. Throughout the paper, the critical trends observed in the field are discussed, including advancements in DA-based hardware architectures. Moreover, the challenges encountered in implementing DA-based AF is also discussed. The continued evolution of DA techniques to cater to the demands of modern AF applications, including real-time processing, resource-constrained environments, and high-dimensional data streams is anticipated. In conclusion, this paper consolidates the current state of applying DA to AF algorithms, offering insights into prevailing trends, discussing challenges, and presenting future research and development in the field. The fusion of these two domains holds promise for achieving improved computational efficiency, reduced hardware complexity, and enhanced performance in various signal processing applications.  
\end{abstract}


\section{Introduction}
\begin{figure}[t]
  \centering
  \includegraphics[width=0.90\linewidth]{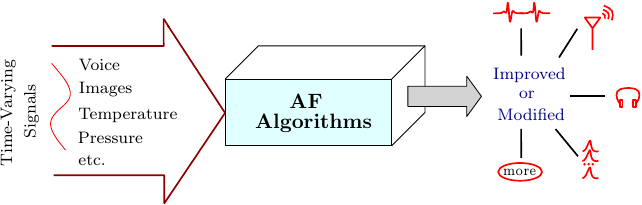}
  \caption{An overview of applications of AF algorithms.}
\label{fig1}
\end{figure}

In the realm of modern signal processing, the extraction of pertinent information from noisy or distorted signals is a ubiquitous challenge. Adaptive filtering (AF) algorithms become essential for addressing this challenge in terms of improvisation across a spectrum of applications \cite{haykin1996estimation,farhang2013adaptive}, including wireless communications \cite{zoltowski2001recent}, image and audio processing \cite{bose2003digital}, biomedical signal analysis \cite{he2004removal}, and more, as shown in Fig. \ref{fig1}. These algorithms dynamically adjust their parameters to accommodate changing signal characteristics, thereby enabling the extraction of valuable information from complex and time-varying data. However, the computational intensity of AF poses a substantial bottleneck in achieving real-time processing, especially in resource-constrained scenarios.

\begin{figure}[t]
  \centering
  \includegraphics[width=0.85\linewidth]{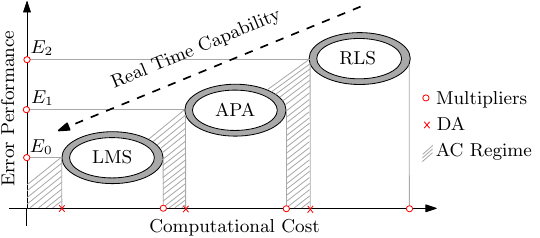}
  \caption{Comparison of error performance, computational cost, and real-time efficiency across LMS, APA, RLS with different implementation approaches such as multipliers-based, DA and AC, where $E_0 < E_1 < E_2$.}
\label{fig2}
\end{figure}

Depending upon the requirements, AF algorithms are broadly classified as least-mean squares (LMS)  \cite{widrow1967adaptive, widrow1977stationary}, affine projection algorithm (APA) \cite{ozeki1984adaptive} and 
recursive-least squares (RLS) \cite{cioffi1984fast}. LMS is an AF algorithm that minimizes mean square error, suitable for slowly varying systems or unknown input signal statistics; it's effective but compromised by rapid changes or noise. It has a low computational cost and good real-time capability due to its simplicity \cite{martin1986adaptive}. APA enhances LMS with multiple regressors for faster convergence and better tracking; it's more complex but applicable in many real-time cases. RLS excels in convergence and tracking, handling changing systems, but demands more resources due to matrix operations, potentially challenging real-time processing in low-resource settings. The choice depends on application needs and the resources available. A typical diagram to illustrate the error performance, computational cost, and real-time capabilities of these algorithms is shown in Fig. \ref{fig2}. The LMS algorithm offers computational simplicity, better real-time capabilities and provides satisfactory error performance as compared to other two algorithms. Due to these reasons, it is often optimized for hardware implementation on application-specific integrated circuits (ASICs) and field-programmable gate arrays (FPGAs) \cite{khan2022asic,di2020low}.    

Approximate computing (AC) in the context of AF introduces a trade-off between real-time capability and error performance \cite{han2013approximate}. By allowing controlled deviations from precise computations, AC techniques aim to reduce the computational cost of AF algorithms. For example, applying AC to the RLS algorithm can yield error performance similar to the APA by reducing the computational load \cite{feng2005fast}. Similarly, applying AC to APA can result in error performance similar to the LMS algorithm, again by lessening the computational demands \cite{garg2016bit}. These transformations enabled by AC play a pivotal role in catering to specific application needs. These approximations can lead to faster processing and improved real-time responsiveness, crucial for applications with stringent timing requirements. However, this advantage comes at the cost of introducing errors in the filtering process, which can impact the overall accuracy of the output signal \cite{jiang2018high}. Striking the right balance between achieving real-time responsiveness and maintaining an acceptable level of error performance is a central challenge for real-world applications.

Distributed arithmetic (DA) stands out as a remarkable approach for AF that offers a distinct advantage over AC. Unlike AC, it provides a means to achieve significant computational efficiency while maintaining error performance at par with exact calculations. This technique capitalizes on pre-computed values and binary representations, resulting in accelerated computations without any compromise on the accuracy of the filtering process. By harnessing DA, the AF algorithms can efficiently process data in real-time scenarios without sacrificing the precision demanded by critical applications, positioning it as an essential tool for achieving both high-performance and high-fidelity outcomes. It was first introduced by Croisier in \cite{croisier1973digital} and later improved by A. Peled and B. Liu in \cite{peled1974new}. Initially, it was applied to second-order infinite impulse response (IIR) filters \cite{white1989applications}. The capabilities of this technique were extended to realizing higher-order IIR filters through strategic combinations. It revolutionized DSP by proposing a novel approach for efficient multiplication using pre-computed partial products stored in a lookup table (LUT) \cite{cowan1981new,cowan1983digital,sicuranza1986adaptive, smith1988nonlinear, tsunekawa2001high, sharma2016alternative}. While initially devised for efficient multiplication in DSP applications, the intersection of DA with AF algorithms has opened up new vistas for optimizing computation and overcoming the limitations of conventional filtering methods. The essence of DA lies in its capacity to exploit the parallelism and eliminate the redundant operations, leading to streamlined multiply-accumulate (MAC) operations that are intrinsic to AF. By capitalizing on the inherent parallelism of DA, it is possible to reduce the computational complexity of AF, thereby accelerating the processing time and enabling real-time implementations \cite{douglas2017introduction,jaggernauth1985real}. Some latest developments in the field of neural networks using DA have been reported in \cite{yalamarthy2019low,khan2022architectural,alhartomi2023low}. 

The motivation behind writing this article on DA-based AF algorithms stems from the growing significance and potential in the field of signal processing. As this technique gains momentum, it is imperative to provide a consolidated resource that explores its advantages, and limitations. By focusing on trends, the article aims to shed light on the latest advancements and emerging practices in the field, guiding researchers and practitioners toward novel solutions. Addressing the challenges associated with DA-based AF algorithms will foster a deeper understanding of their practical feasibility and potential limitations. Overall, this article seeks to serve as a pivotal resource that not only captures the current state of the field but also paves the way for its future growth and innovation. In the subsequent section, the discussion of DA-based LMS algorithms are presented, laying the groundwork for a deeper understanding of their fusion.

\section{Literature review of DA-based architecture for LMS and its variants}
Due to the complexity limitations of APA and RLS algorithms, DA is predominantly applied to a standard LMS and its high-speed variants. In the future, there is potential to combine DA with AC techniques, enabling the approximate implementation of APA and RLS AFs. This convergence could yield comparable error performance, computational efficiency, and real-time capabilities to that of LMS AFs. Usually, AFs represent robust self-designing systems utilized across a diverse range of DSP applications including active power filters \cite{pereira2011new}, bearing prognosis \cite{li2019prognosis}, active noise control \cite{chang2023complete, akhtar2023developing}, and 5G communication systems \cite{ khan2020high,khan2017energy,khan2016low,khan2016efficient}. Among different AF algorithms, finite impulse response (FIR) based structures stand out due to their inherent stability characteristic. It comprises an adjustable coefficient FIR filter alongside a component responsible for updating these coefficients through adaptive algorithms.

DA presents a computationally efficient technique for bit-serial inner product computations (IPCs) in AFs \cite{allred2003implementation,allred2005lms,guo2011two,prakash2013low,park2013low,meher2011high,khan2018optimal,khan2018improved,khan2022high} and matrix-vector multiplications in neural networks \cite{yalamarthy2019low,khan2022architectural,alhartomi2023low}. It proves valuable in implementing digital filtering and machine learning processes without necessitating hardware multipliers. By pre-computing and storing combinations of filter coefficients within memory elements, typically look-up tables (LUTs), the method allows for their retrieval using address formation based on input sample bits or coefficients, from the least-significant bit (LSB) to the most-significant bit (MSB). The LUT outputs are then subjected to shifting and accumulation to compute the final digital filter output. Nonetheless, the DA method faces a major drawback in the form of exponential growth in the number of partial products stored within LUTs, particularly for large filter orders. Complications also emerge when employing the DA approach to implement AFs. The dynamic nature of AFs, which demands the updating of filter coefficients at every iteration, renders pre-computing filter coefficients unfeasible. This challenge significantly diminishes the benefits of the DA method. In the following, the existing DA-based architectures for implementing LMS AF and its high-speed variants are discussed. 
\subsection{DA-based LMS AFs}
The work presented in \cite{allred2005lms} for DA-based LMS AF has captured the interest of researchers, driving them to explore architectural innovations aimed at achieving optimized area, power, and throughput designs for LMS AFs \cite{sarma2019novel}. They introduced an effective update technique employing two distinct LUTs: a filtering LUT and an auxiliary LUT. The update process for the auxiliary LUT entails re-purposing the upper half of the previous LUT's contents and adding the most recent sample to calculate the remaining half. Meanwhile, the filtering LUT is updated by scaling and integrating the contents of the auxiliary LUT into corresponding addresses, in line with the LMS coefficient update protocol. It is noteworthy that this update approach does not impose any assumptions on the statistical characteristics of the input signal. An improved version of \cite{allred2005lms} has been presented in \cite{khan2017low}. Subsequently, a novel conjugate DA structure was proposed for AFs \cite{huang2006conjugate}. In contrast to the conventional or direct DA structure, the conjugate DA employs AF coefficients to formulate addresses for the filtering LUT. This technique alleviates the necessity for an auxiliary LUT, effectively halving the required memory and enhancing throughput. Later on, two separate designs for two's complement (TC) and offset-binary coding (OBC) schemes, utilizing the conjugate DA approach, were independently introduced \cite{guo2011two}. For TC DA based AFs, the LUT update method follows the procedure described in \cite{allred2005lms}. In the case of OBC DA based AFs, a distinct update mechanism is adopted. This involves subtracting a term computed using the most recent and the oldest samples from the LUT content at even addresses and adding it to the content at odd addresses to compute the succeeding iteration's second and first halves of the LUT content, respectively. The conjugate DA structure was also extended to implement sliding block (SB) AFs \cite{jones1993efficient} and its modified version of SB-DA AFs \cite{huang2011modified}. In SB-DA AFs, the processing of input samples by the filter evolves slowly, with only one sample changing per iteration. Through the utilization of the LUT decomposition method established in \cite{wei1986multimemory,chiu1987realization}, only one of the smaller LUTs, containing the oldest sample, requires modification, while an additional LUT is designated to accommodate the newest sample. To execute such an operation while reusing the LUT contents, circular shifting of the coefficients becomes essential. This concept of LUT reusability is extensively employed in the implementation of block-processing DA-AFs. 
In \cite{prakash2013low}, the framework from \cite{allred2005lms} was adopted while leveraging the OBC scheme. A novel update for the auxiliary LUT was introduced, enabling its division into two smaller LUTs based on even and odd address locations. An improved version of \cite{prakash2013low} has been presented in \cite{khan2017new}. This approach facilitated the development of a low-area and high-throughput design for LMS AF. Further reduction in area has been achieved without splitting the LUTs into even and odd components, as demonstrated in \cite{khan2018area}. In order to improve the throughput, two high-throughput designs based on TC and OBC schemes have been recently presented in \cite{khan2022two}, utilizing the frameworks of \cite{guo2011two,prakash2013low}. 

In the following, two high-speed variants of LMS algorithm, namely, Delayed LMS (DLMS) \cite{van2001efficient} and Block LMS (BLMS) \cite{farhang2000analysis} will be discussed. In DLMS, delay registers are introduced in the adaptation loop, while in BLMS, block inputs are processed in parallel. In the case of the BLMS algorithm, the frequency domain variant is popular as it allows computational savings in hardware. Few fast Fourier Transform (FFT) \cite{hazarika2023efficient,hazarika2020area,hazarika2019energy,hazarika2019high,hazarika2019low} based BLMS AFs are reported in \cite{srivastava2021fast,khan2023analyzing}. Since both variants of LMS are capable of providing higher speeds than standalone versions, it was intriguing to explore their DA-based implementations.  

\begin{table}[]
\centering
\caption{Different DA-based architecture for LMS and its variants}
\begin{threeparttable}
\begin{tabular}{|l|l|l|l|}
\hline
{\bf{Algo.}}         & {\bf{Design}} & {\bf{Type}} & {\bf{No. of LUTs}}, {\bf{LUT-Type}} \\ \hline
\multirow{6}{*}{{\bf{LMS}}} &  Allred et al. \cite{allred2005lms}      &    TC  & Two, SRAM    \\ \cline{2-4}
& Guo et al. \cite{guo2011two}       &    TC  & One,  SRAM    \\ \cline{2-4}
& Guo et al. \cite{guo2011two}       &    OBC  & One, ($\times \frac{1}{2}$) SRAM    \\ \cline{2-4}
& Surya et al. \cite{prakash2013low}      &    OBC  & Four ($\times \frac{1}{4}$) SRAM    \\ \cline{2-4}
& Khan et al. \cite{khan2022two}       &    TC  & Two,  ($\times \frac{1}{2}$) SRAM    \\ \cline{2-4}
& Khan et al. \cite{khan2022two}       &    OBC  & Two ($\times \frac{1}{4}$) SRAM    \\ \cline{1-4}
\multirow{4}{*}{{\bf{DLMS}}} & Meher et al. \cite{meher2011high}       &  TC    & One, Hardware    \\ \cline{2-4} 
&    Park et al. \cite{park2013low}   &    TC  &  One, Hardware$^{*}$   \\ \cline{2-4}
&   Khan et al. \cite{khan2018optimal}  &  OBC     &  One, Hardware$^{**}$  \\ \cline{2-4}
&   Khan et al. \cite{khan2022high}  &  TC     &  One, Hardware$^{***}$   \\ \cline{1-4}

\multirow{6}{*}{{\bf{BLMS}}} &    Basanty et al. \cite{mohanty2012high}   &  TC    &  $L$, SRAM (Shared)     \\ \cline{2-4} 
& Basanty et al. \cite{mohanty2015lut}       & TC      &  $L$, Hardware (Shared)   \\ \cline{2-4}
& Khan et al. \cite{khan2018analysis}      & OBC$^{\dag}$      &  $L$, SRAM (Shared)   \\ \cline{2-4}
& Khan et al. \cite{khan2020high}      & OBC      &  $L$, SRAM (Shared)   \\ \cline{2-4}
& Khan et al. \cite{khan2020partial}      & TC      &  $L$, Hardware (Shared, Partial)   \\ \cline{2-4}
& Khan et al. \cite{khan2021efficient}      & OBC      &  $L$, Hardware (Shared, Partial)   \\ \cline{1-4}
\end{tabular}
\begin{tablenotes}
      \item 
$^{\dag}:$ Recursive OBC, $\times \frac{1}{2}$: LUT size reduced by half compared to \cite{allred2005lms}, $\times\frac{1}{4}$: LUT size reduced to one-fourth of \cite{allred2005lms},
$^{*}:$ Represents a more hardware-optimized version of \cite{meher2011high}. $^{**}:$ Represents a more hardware-optimized version of \cite{park2013low}. $^{***}:$ Represents a more hardware-optimized version of \cite{khan2018optimal}. Shared: Utilized for both the computation of filter output and coefficient increment term. Partial: Stores partial LUT contents.
      \end{tablenotes}
       \end{threeparttable}
 
\end{table}

\subsection{DA-based DLMS AFs}
To improve the processing speed of DA-AF designs, pipelining delays have been employed in the error feedback path for coefficient adaptation \cite{wallace1964suggestion,long1989lms,meher2011high,park2013low,khan2018optimal,khan2017vlsi,khan2022high,arenas2006mean}. It serves to reduce critical-path constraints through re-timing of delays, and thereby improves the throughput of the system. The first DA-based DLMS AF design was presented in \cite{meher2011high}, based on the TC DA. Subsequently, a pipelined AF using TC conjugate DA structure was introduced in \cite{park2013low}. It is more hardware optimized version of \cite{meher2011high} due to the use of hardware LUTs, resulting the savings in area and power. More importantly, this also eliminates the LUT access time issues, and allows the possibility of sharing partial products. In DA-based designs, pipelining introduces delay elements that separate various processing stages, and accompanying register-based LUTs to eliminate access time issues. Moreover, pipelined registers are introduced between filtering and weight update processes \cite{meher2011high,park2013low}. Register-based LUTs perform parallel and online computation of DA combinations, albeit at the expense of an exponential increase in the required number of registers. Addressing this issue, proposals for multiplexed LUT and AND cell LUT configurations were put forth in \cite{khan2017vlsi}. Enhanced performance was achieved through an improved DA-AF design in \cite{khan2022high}, where an AND cell LUT served as the base unit LUT, and delay elements were integrated into the adder tree. Due to its advantages, it has been extended to a convex combination of two LMS for convergence enhancement in \cite{arenas2006mean,khan2018enhanced}. This approach replaces two combined AFs with a single DA-AF unit featuring two multiplexed step sizes \cite{khan2018improved}. 

\subsection{DA-based BLMS AFs}
The realm of block processing designs has been extensively explored in studies such as \cite{baghel2011fpga,mohanty2012high,mohanty2015lut,khan2018analysis,khan2020high,khan2020partial,khan2021efficient}. In these designs, with a given block length denoted as $L$, block digital filters manage $L$ input vectors and compute an equivalent number of outputs in a single iteration. Consequently, opting for a larger value of $L$ yields a proportional increase in hardware costs, ultimately leading to a correspondingly elevated throughput. Within block AFs, the DA methodology has been harnessed, as exemplified in \cite{baghel2011fpga}, where it finds application in efficiently implementing the FFT as part of the Fast Block LMS method. Notably, the inaugural formulation of Block LMS (BLMS) through a DA framework, built upon the conjugate TC DA-based structure, was accomplished in \cite{mohanty2012high}. Here, a comparable approach to that of Sliding Block DA (SB-DA) was adopted for the updating of LUT contents. A conglomerate of LUTs, designated as a processing element, where the oldest samples reside, undergoes overwriting through combinations of the most recent samples during each iteration. It is termed "inter-iteration LUT reuse," the technique involves the recycling of LUT contents between successive iterations, specifically within a processing element. The remaining processing elements, housing intermediate samples, remain unaffected. The implementation of this approach relies on a register array-based LUT, which facilitates the sharing of LUT contents during a single iteration. This practice, known as "intra-iteration LUT sharing," has been optimized in \cite{mohanty2015lut}. This optimization benefits from the fact that successive LUTs within a processing element share three overlapping samples, allowing half of the previous LUT's contents to be shared with the subsequent LUT. As a result, this strategy diminishes memory complexity. A corresponding OBC DA-based implementation was introduced in \cite{khan2018analysis}, refining the initial design proposed in \cite{mohanty2015lut} by incorporating an even-odd LUT decomposition utilizing the OBC DA scheme \cite{khan2020high}. Subsequent advancements embraced intra-iteration adaptations for the TC DA in \cite{khan2020partial}, further refined in \cite{khan2021efficient} based on OBC DA scheme.

\section{Formulation of LMS AF using DA}
\begin{figure}[t]
  \centering
  \includegraphics[width=0.58\linewidth]{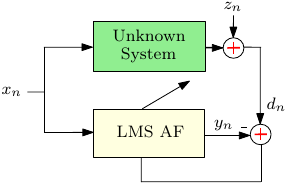}
  \caption{Block diagram of LMS AF with a system identification problem.}
\label{fig56}
\end{figure}

At the top-level, LMS AF for a system identification problem is shown in Fig. \ref{fig56}. In this section, the LMS AF is formulated using TC or OBC DA. The analysis can be extended similarly for DLMS and BLMS algorithms. In this context, the output $y_n$ of an $N^{\textnormal{th}}$ order LMS AF using an IPC at time instant $n$ can be defined as
\begin{equation}\label{eq1}
y_n={\boldsymbol{w}}^{T}_n{\boldsymbol{x}}_n=\sum\limits_{i=0}^{N-1}w_{n}(i)x_{n-i}
\end{equation}
Here, ${\boldsymbol{x}}_n=[x_{n}, x_{n-1},..., x_{n-N+1}]^{T}$ signifies the input vector, while ${\boldsymbol{w}}_n=[w_{n}(0), w_{n}(1),..., w_{n}({N-1})]^{T}$ represents the coefficient vector. Subsequently, an error term $e_n$ is calculated by subtracting the obtained output $y_n$ from the desired signal $d_n$, leading to the equation:
\begin{equation}\label{eq:error}
e_n=d_n+z_n-y_n
\end{equation}
In this scenario, $z_n$ represents the additive white Gaussian noise observed at the output of an unknown system, as illustrated in Fig. \ref{fig56}. The aim is to adjust the coefficients during each iteration according to the computed error outlined in (\ref{eq:error}), according to 
\begin{equation}\label{eq:dblms}
{\boldsymbol{w}}_{n+1}={\boldsymbol{w}}_{n}+\mu e_n{\boldsymbol{x}}_n
\end{equation}
This article explores the possibilities of DA-based realization of LMS AF. To begin with formulation of IPC given in (1) and coefficient update equation given in (3), it is important understand whether inputs or coefficients are represented in TC/OBC form. For instance, the work reported in \cite{allred2005lms} has represented inputs in TC form, while the work presented in \cite{guo2011two} has represented the coefficients in TC/OBC form, and so on. For clarity, the coefficients are represented in TC/OBC form in this work, however one can also consider the inputs to be in TC/OBC form.     

\subsection{IPC using TC DA}
This subsection delves into the methodology of IPC using TC DA. The foundation of this design is the representation of filter coefficients, denoted as $w_n(i)$, in a signed TC $B$-bit format as
\begin{equation}\label{eq4}
w_n({i})=-w_n({i,0})+\sum\limits_{j=1}^{B-1}w_n({i,j})2^{-j}
\end{equation}
where $i\in [0, N-1]$. Substituting (\ref{eq4}) into (1), and after some simplification, yields:
\begin{equation}\label{eq5}
y_n=\sum\limits_{j=0}^{B-1}a_n({j})2^{-j}
\end{equation}
with
\begin{equation}\label{eq6}
a_n({j})=(-1)^{\lfloor {(B-1-j)}/{(B-1)} \rfloor}b_n({j})
\end{equation}
where $\lfloor \boldsymbol{\cdot} \rfloor$ represents the greatest integer function, and $j$ is confined within the range $[0, B-1]$. The resulting pattern shows that the filter partial products $a_n(j)$ undergo a shift-accumulation over $B$ clock cycles, with a sign inversion occurring during the $0^{\textnormal{th}}$ clock cycle. The term $b_n({j})$ is further defined by 
\begin{equation}\label{eq7}
b_n({j})=\sum\limits_{i=0}^{N-1}x_{n-i}w_n({i,j})
\end{equation}
In this equation, $w_n({i,j})$ represents the bit-slices of the filter coefficients, with values in the range $[0,1]$. This allows for the representation of $b_n({j})$ through $2^N$ binary combinations, which can be pre-computed and stored in a LUT, as illustrated in Fig. 4(a). Note that the size of the LUT grows exponentially with $N$, which becomes problematic for higher values of $N$. This issue could be mitigated by encoding the coefficients and incorporating bit-symmetries using OBC scheme. 
\subsection{IPC using OBC DA}
In this methodology, the encoding of filter coefficients takes the form: $w_{i}=\frac{1}{2}[w_{i}-{\overline{w}}_{i}]$, where ${\overline{w}}_{i}$ represents the TC of $w_{i}$. Building upon (\ref{eq4}), it can be expanded further as 
\begin{equation}\label{eq9}
w_{i}=\frac{1}{2}\bigg[-(w_{i,0}-{\overline{w}}_{i,0})+\sum\limits_{j=1}^{B-1}(w_{i,j}-{\overline{w}}_{i,j})2^{-j}-2^{-(B-1)}\bigg]
\end{equation}
Here, $\Delta w_{i,j}=(w_{i,j}-{\overline{w}}_{i,j})$ signifies the disparity between the bit-slices of filter coefficients and their corresponding TCs. Substituting (8) into (\ref{eq1}) and undergoing subsequent simplification yields 
\begin{align}\label{eq10}
y_n &=\sum\limits_{j=0}^{B-1}\bigg(\frac{1}{2}\sum\limits_{i=0}^{N-1}x_{n-i}c_{i,j}\bigg)2^{-j}-\bigg(\frac{1}{2}\sum\limits_{i=0}^{N-1}x_{n-i}\bigg)2^{-(B-1)}
\end{align}
with 
\begin{equation}
c_{i,j} = (-1)^{\lfloor {(B-1-j)}/{(B-1)} \rfloor} \Delta w_{i,j}    
\end{equation}
The equation (9) is then transformed into a more concise form as represented as 
\begin{equation}\label{eq11}
y_n=\sum\limits_{j=0}^{B-1}d_{j}2^{-j}+d_{initial}2^{-(B-1)}
\end{equation}
where the term $d_{j}$ and $d_{initial}$, are respectively, expressed as
\begin{equation}\label{eq12}
d_{j}=\frac{1}{2}\sum\limits_{i=0}^{N-1}x_{n-i}c_{i,j},  \qquad
d_{initial}=-\frac{1}{2}\sum\limits_{i=0}^{N-1}x_{n-i}
\end{equation}
It is noteworthy that the term $d_{j}$ has $2^{N}$ potential binary combinations derived from input samples, which can be pre-computed and stored in a LUT akin to the Type I structure. However, the nature of these stored contents differs from that of the Type I structure, as $c_{i,j}$ values exist within the range of $[-1,1]$. Based on the value of $c_{N-1,j}$, it is apparent that the contents present at the first half and second half address locations of the LUT are TC of each other. Therefore, it is sufficient to store only $2^{N-1}$ half combinations in the LUT, as illustrated in Fig. 4(b).

\begin{figure}[t]
  \centering
  \includegraphics[width=1\linewidth]{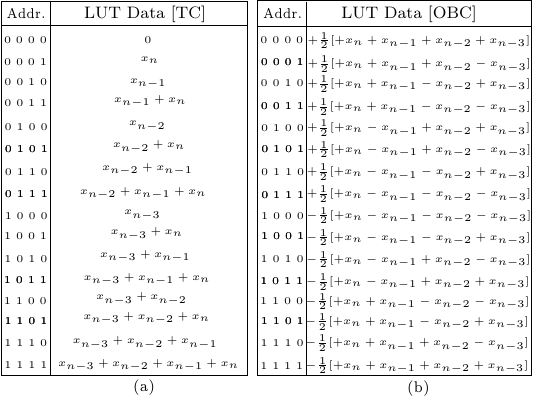}
  \caption{LUT contents of an IPC with (a) TC DA and (b) OBC DA for $N=4$.}
\label{fig4}
\end{figure}
\subsection{Coefficient Update using TC and OBC DA}
By representing the coefficients ${\boldsymbol{w}}_{n}$ in signed $B$-bit TC form, we have
\begin{equation}\label{eq53}
{\boldsymbol{w}}_{n}={\boldsymbol{W}}_{n}{\boldsymbol{s}}
\end{equation}
where ${\boldsymbol{s}}=[-2^{0}, 2^{-1},..., 2^{-(B-1)}]$ is the vector of binary weights and ${\boldsymbol{W}}_{n}=[w_n({i,j})]_{N\times B}$ is the coefficient bit-slice matrix in TC form, where $w_n({i,j})\in [0, 1]$ $\forall$ $0 \leq  i \leq N-1$ and $0 \leq  j \leq B-1$. Substituting (13) into (3), we have 
\begin{equation}\label{eq51}
{\boldsymbol{W}}_{n+1}{\boldsymbol{s}}={\boldsymbol{W}}_{n}{\boldsymbol{s}}+\mu e_n {\boldsymbol{x}}_n
\end{equation}
It is clear from (14) that LMS algorithm using TC DA is same as standard LMS except that the coefficients are sliced into bits after update process. Unlike (14), OBC exploits the bit-symmetries for representation of coefficients in OBC form. By re-writing the coefficients as ${\boldsymbol{w}}_n=\frac{1}{2}[{\boldsymbol{w}}_n-(-{\boldsymbol{w}}_n)]=\frac{1}{2}[{\boldsymbol{w}}_n-{\overline{\boldsymbol{w}}}_n-1]$, where ${\overline{\boldsymbol{w}}}_n$ is the 1's complement of ${\boldsymbol{w}}_n$. Therefore, we can express the coefficients ${\boldsymbol{w}}_n$ using (14) in OBC form as
\begin{equation}
{\boldsymbol{w}}_{n}={\boldsymbol{W}}_{n}{\boldsymbol{s}}=\frac{1}{2}\left[(\tilde{\boldsymbol{W}}_{n}-1){\boldsymbol{s}}\right]
\end{equation}
where $\tilde{\boldsymbol{W}}_{n}=[\tilde w_n({i,j})]_{N\times B}$ $\forall$ is the coefficient difference bit-slice matrix in OBC form with $\tilde{w}_n({i,j})\in [-1,1]$. Substituting (15) into (3), and after some simplification, we have 
\begin{equation}\label{eq52}
\tilde{\boldsymbol{W}}_{n+1}{\boldsymbol{s}}=\tilde{\boldsymbol{W}}_{n}{\boldsymbol{s}}+2\mu e_n {\boldsymbol{x}}_n
\end{equation}
Clearly, the coefficient update equation is affected by the bit-symmetries i.e., it involves extra scaling of $2\times$ alongwith the step-size. 

\section{Trends, Challenges, and Future Prospects}
In this section, the trends, challenges and future prospects of DA-based implementation for LMS algorithm and variants will be discussed. 
\subsection{Trends}
DA is an efficient technique used for the implementation of LMS AFs, involving IPCs of binary vectors with shift and add operations. This aims to reduce computation complexity and resource requirements. This could also involve research into optimized hardware architectures, algorithmic enhancements, and novel techniques for accelerating DA operations. Due to high computational efficiency, researchers might explore hybrid approaches that combine DA or AC with other adaptive filtering techniques to improve overall performance and flexibility. With the increasing demand for real-time signal processing in various fields such as communications, audio processing, and image processing, DA-based AF may have been adapted and optimized for low-latency applications. Energy-efficient signal processing is a significant consideration in many applications. Researchers might work on optimizing DA-based adaptive filters for reduced power consumption, making them suitable for battery-operated devices and energy-constrained environments. Given the trend toward parallel and distributed computing architectures, researchers may have explored ways to parallelize and distribute the computations involved in DA-based AF. This could lead to improved performance and scalability. Researchers might explore ways to integrate machine learning techniques with DA-based AF to create hybrid adaptive systems that can learn from data and adapt more effectively to changing conditions. In recent works on DA-based implementation, hardware LUT (or sometimes it is referred to as LUT-less design) removes the bottleneck of LUT access time on the performance of LMS AFs, as mentioned before. In the latest development on high-speed implementation of LMS algorithm, a comprehensive analysis that combines both pipelining and block processing has been presented in \cite{khan2023stochastic}. This work has developed a stochastic model which established the relation of step-size in terms filter order and delay/block size (or speedup).   

\subsection{Challenges}
DA-based AF, while a powerful technique, also comes with certain challenges and considerations that need to be addressed. Since DA relies on binary representations and shift-and-add operations, which can lead to limited precision and accumulation of errors over time. This can impact the overall accuracy and performance of the filter, especially in applications requiring high precision. Although applying DA to AF was seem to be initially difficult as they are also involve coefficient adaptation. The issue of auxillary LUTs for DA-based implementation was overcome wth conjungate DA. For real-time implementation, choosing appropriate adaptation algorithms and strategies that converge quickly while maintaining stability is essential. Inherently, DA-based AFs are linear and may struggle when dealing with non-linear or time-varying systems. Adapting to changes in system behavior while maintaining filter stability can be difficult. While DA can reduce the number of multiplications in several AF implementations, it can lead to increased complexity in terms of shift-and-add operations, especially for higher input or coefficient precision. This may pose challenges in terms of hardware implementation, especially for high-speed. Previously, DA requires careful management of memory resources for storing coefficients and intermediate results. This has been replaced hardware LUTs which are important for real-time or embedded systems. The filter length and step-size value play a crucial role in determining its performance and complexity. Striking a balance between achieving the desired filtering characteristics and managing computational demands can be challenging. In addition, step-size can also adaptation rate, for instance, achieving fast adaptation speed while ensuring stability can be a delicate balance. Aggressive adaptation can lead to instability, while overly cautious adaptation might result in slower convergence. While DA has inherent parallelism especially when realizing the partial products with hardware LUTs, efficiently parallelizing the computations across multiple processing units or cores can be challenging, especially for complex filter structures. It's important to note that researchers continue to work on addressing these challenges through algorithmic improvements, hardware innovations, and hybrid approaches that combine DA with other techniques. 

\subsection{Future}
The future research could focus on designing more efficient and specialized hardware architectures optimized for DA-based AF. These architectures might leverage emerging technologies such as FPGAs, ASICs, or even neuromorphic computing for improved performance and energy efficiency. Furthermore, combining DA-based AF with other advanced filtering techniques, such as machine learning and neural networks, could lead to hybrid adaptive systems that exploit the strengths of each approach. This might involve using machine learning to initialize or fine-tune DA filters, enabling more adaptable and robust filtering in complex scenarios.
Researchers may also work on extending the capabilities of DA-based filters to handle non-linear and time-varying systems more effectively. Developing adaptive strategies that can dynamically adjust filter parameters to accommodate changing system behaviors could open up new applications. As the demand for real-time processing continues to grow in applications like 6G communications, autonomous systems, and sensor networks, the future scope of DA-based AF could involve optimizing the technique for low-latency and high-throughput processing. Specifically, high-speed LMS recently developed in \cite{khan2023stochastic} can be deployed in 6G communication system due to both pipelining and block processing. With the proliferation of edge computing and IoT devices, there could be increased interest in developing lightweight and energy-efficient DA-based AF that can operate on resource-constrained platforms. As technology unfolds, new fields may emerge that can benefit from DA-based AF. For example, applications in quantum computing, bio-informatics, or novel communication systems might present opportunities for the application of this technique.

\section{Conclusion}
In summary, the utilization of DA-based AF algorithms has garnered considerable attention in recent years, owing to its potential to bolster computational efficiency and diminish resource requirements. This paper has delved into the application of DA in AF algorithms, scrutinizing trends, addressing challenges, and delineating future prospects. It commenced by furnishing an overview of both DA and AF algorithms, elucidating their respective merits and established applications. The subsequent exploration focused on integrating DA into AF algorithms, demonstrating its efficacy in optimizing multiply-accumulate operations and alleviating the computational burden associated with AF algorithms. Although the majority of designs center on the implementation of LMS AFs due to their simplicity and commendable performance, there exists a dearth of DA-based AF designs addressing more advanced adaptive algorithms such as APA and RLS.
Notably, a recurring trend has surfaced, wherein various design types ultimately gravitate towards conjugate OBC/TC DA based implementation with hardware LUTs. Such designs obviate the need for auxiliary LUTs and further reduce memory access issue. In a broader context, it is discernible that initial structures predominantly embrace direct TC DA based paradigms, subsequently undergoing refinement via the integration of conjugate DA or OBC DA-based methodologies. An additional promising direction for future research revolves around the strategic pursuit of novel adaptive algorithms such as presented in \cite{khan2023stochastic} amenable to the DA implementation framework. To summarize, this paper consolidates the current state of applying DA to AF algorithms, providing insights into prevailing trends, addressing challenges, and outlining future research and development in the field. The fusion of these two domains holds promise for achieving improved computational efficiency, reduced hardware complexity, and enhanced performance across various signal processing applications.

\bibliographystyle{IEEEtran}
\bibliography{IEEEabrv,main}

\begin{thebibliography}{10}
\providecommand{\url}[1]{#1}
\csname url@samestyle\endcsname
\providecommand{\newblock}{\relax}
\providecommand{\bibinfo}[2]{#2}
\providecommand{\BIBentrySTDinterwordspacing}{\spaceskip=0pt\relax}
\providecommand{\BIBentryALTinterwordstretchfactor}{4}
\providecommand{\BIBentryALTinterwordspacing}{\spaceskip=\fontdimen2\font plus
\BIBentryALTinterwordstretchfactor\fontdimen3\font minus
  \fontdimen4\font\relax}
\providecommand{\BIBforeignlanguage}[2]{{%
\expandafter\ifx\csname l@#1\endcsname\relax
\typeout{** WARNING: IEEEtran.bst: No hyphenation pattern has been}%
\typeout{** loaded for the language `#1'. Using the pattern for}%
\typeout{** the default language instead.}%
\else
\language=\csname l@#1\endcsname
\fi
#2}}
\providecommand{\BIBdecl}{\relax}
\BIBdecl

\bibitem{haykin1996estimation}
S.~Haykin, ``Estimation theory,'' \emph{Adaptive Filter Theory, Upper Saddle
  River, NJ: Prentice-Hall}, pp. 899--904, 1996.

\bibitem{farhang2013adaptive}
B.~Farhang-Boroujeny, \emph{Adaptive filters: theory and applications}.\hskip
  1em plus 0.5em minus 0.4em\relax John Wiley \& Sons, 2013.

\bibitem{zoltowski2001recent}
M.~D. Zoltowski, M.~Joham, and S.~Chowdhury, ``Recent advances in reduced-rank
  adaptive filtering with application to high-speed wireless communications,''
  \emph{Digital Wireless Communication III}, vol. 4395, pp. 1--15, 2001.

\bibitem{bose2003digital}
T.~Bose and F.~Meyer, \emph{Digital signal and image processing}.\hskip 1em
  plus 0.5em minus 0.4em\relax John Wiley \& Sons, Inc., 2003.

\bibitem{he2004removal}
P.~He, G.~Wilson, and C.~Russell, ``Removal of ocular artifacts from
  electro-encephalogram by adaptive filtering,'' \emph{Medical and biological
  engineering and computing}, vol.~42, pp. 407--412, 2004.

\bibitem{widrow1967adaptive}
B.~Widrow, P.~Mantey, L.~Griffiths, and B.~Goode, ``Adaptive antenna systems,''
  \emph{Proceedings of the IEEE}, vol.~55, no.~12, pp. 2143--2159, 1967.

\bibitem{widrow1977stationary}
B.~Widrow, J.~McCool, M.~G. Larimore, and C.~R. Johnson, ``Stationary and
  nonstationary learning characteristics of the lms adaptive filter,'' in
  \emph{Aspects of signal processing}.\hskip 1em plus 0.5em minus 0.4em\relax
  Springer, 1977, pp. 355--393.

\bibitem{ozeki1984adaptive}
K.~Ozeki and T.~Umeda, ``An adaptive filtering algorithm using an orthogonal
  projection to an affine subspace and its properties,'' \emph{Electronics and
  Communications in Japan (Part I: Communications)}, vol.~67, no.~5, pp.
  19--27, 1984.

\bibitem{cioffi1984fast}
J.~Cioffi and T.~Kailath, ``Fast, recursive-least-squares transversal filters
  for adaptive filtering,'' \emph{IEEE Transactions on Acoustics, Speech, and
  Signal Processing}, vol.~32, no.~2, pp. 304--337, 1984.

\bibitem{martin1986adaptive}
K.~W. Martin and M.~T. Sun, ``Adaptive filters suitable for real-time spectral
  analysis,'' \emph{IEEE journal of solid-state circuits}, vol.~21, no.~1, pp.
  108--119, 1986.

\bibitem{khan2022asic}
M.~T. Khan and O.~Gustafsson, ``Asic implementation trade-offs for high-speed
  lms and block lms adaptive filters,'' in \emph{2022 IEEE 65th International
  Midwest Symposium on Circuits and Systems (MWSCAS)}.\hskip 1em plus 0.5em
  minus 0.4em\relax IEEE, 2022, pp. 1--4.

\bibitem{di2020low}
G.~Di~Meo, D.~De~Caro, E.~Napoli, N.~Petra, and A.~G. Strollo, ``Low-power
  implementation of lms adaptive filters using scalable rounding,'' in
  \emph{2020 27th IEEE international conference on electronics, circuits and
  systems (ICECS)}.\hskip 1em plus 0.5em minus 0.4em\relax IEEE, 2020, pp.
  1--4.

\bibitem{han2013approximate}
J.~Han and M.~Orshansky, ``Approximate computing: An emerging paradigm for
  energy-efficient design,'' in \emph{2013 18th IEEE European Test Symposium
  (ETS)}.\hskip 1em plus 0.5em minus 0.4em\relax IEEE, 2013, pp. 1--6.

\bibitem{feng2005fast}
D.-Z. Feng and W.~X. Zheng, ``Fast rls-type algorithm for unbiased
  equation-error adaptive iir filtering based on approximate inverse-power
  iteration,'' \emph{IEEE transactions on signal processing}, vol.~53, no.~11,
  pp. 4169--4185, 2005.

\bibitem{garg2016bit}
B.~Garg, S.~Dutt, and G.~Sharma, ``Bit-width-aware constant-delay run-time
  accuracy programmable adder for error-resilient applications,''
  \emph{Microelectronics Journal}, vol.~50, pp. 1--7, 2016.

\bibitem{jiang2018high}
H.~Jiang, L.~Liu, P.~P. Jonker, D.~G. Elliott, F.~Lombardi, and J.~Han, ``A
  high-performance and energy-efficient fir adaptive filter using approximate
  distributed arithmetic circuits,'' \emph{IEEE Transactions on Circuits and
  Systems I: Regular Papers}, vol.~66, no.~1, pp. 313--326, 2018.

\bibitem{croisier1973digital}
A.~Croisier, D.~Esteban, M.~Levilion, and V.~Rizo, ``Digital filter for pcm
  encoded signals,” us patent no. 3777130,'' 1973.

\bibitem{peled1974new}
A.~Peled and B.~Liu, ``A new hardware realization of digital filters,''
  \emph{IEEE transactions on acoustics, speech, and signal processing},
  vol.~22, no.~6, pp. 456--462, 1974.

\bibitem{white1989applications}
S.~A. White, ``Applications of distributed arithmetic to digital signal
  processing: A tutorial review,'' \emph{IEEE Assp Magazine}, vol.~6, no.~3,
  pp. 4--19, 1989.

\bibitem{cowan1981new}
C.~Cowan and J.~Mavor, ``New digital adaptive-filter implementation using
  distributed-arithmetic techniques,'' in \emph{IEE Proceedings F
  (Communications, Radar and Signal Processing)}, vol. 128, no.~4.\hskip 1em
  plus 0.5em minus 0.4em\relax IET, 1981, pp. 225--230.

\bibitem{cowan1983digital}
C.~Cowan, S.~Smith, and J.~Elliott, ``A digital adaptive filter using a
  memory-accumulator architecture: Theory and realization,'' \emph{IEEE
  Transactions on Acoustics, Speech, and Signal Processing}, vol.~31, no.~3,
  pp. 541--549, 1983.

\bibitem{sicuranza1986adaptive}
G.~Sicuranza and G.~Ramponi, ``Adaptive nonlinear digital filters using
  distributed arithmetic,'' \emph{IEEE transactions on acoustics, speech, and
  signal processing}, vol.~34, no.~3, pp. 518--526, 1986.

\bibitem{smith1988nonlinear}
M.~Smith, C.~Cowan, and P.~Adams, ``Nonlinear echo cancellers based on
  transpose distributed arithmetic,'' \emph{IEEE transactions on circuits and
  systems}, vol.~35, no.~1, pp. 6--18, 1988.

\bibitem{tsunekawa2001high}
Y.~Tsunekawa, K.~Takahashi, S.~Toyoda, and M.~Miura, ``High-performance vlsi
  architecture of multiplierless lms adaptive filters using distributed
  arithmetic,'' \emph{Electronics and Communications in Japan (Part III:
  Fundamental Electronic Science)}, vol.~84, no.~5, pp. 1--12, 2001.

\bibitem{sharma2016alternative}
P.~K. Sharma, M.~T. Khan, and S.~R. Ahamed, ``An alternative approach to design
  reconfigurable mixed signal vlsi da based fir filter,'' in \emph{2016 IEEE
  Students’ Technology Symposium (TechSym)}.\hskip 1em plus 0.5em minus
  0.4em\relax IEEE, 2016, pp. 284--288.

\bibitem{douglas2017introduction}
S.~C. Douglas, ``Introduction to adaptive filters,'' in \emph{Digital signal
  processing fundamentals}.\hskip 1em plus 0.5em minus 0.4em\relax CRC Press,
  2017, pp. 467--484.

\bibitem{jaggernauth1985real}
J.~Jaggernauth, A.~Loui, and A.~Venetsanopoulos, ``Real-time image processing
  by distributed arithmetic implementation of two-dimensional digital
  filters,'' \emph{IEEE transactions on acoustics, speech, and signal
  processing}, vol.~33, no.~6, pp. 1546--1555, 1985.

\bibitem{yalamarthy2019low}
K.~P. Yalamarthy, S.~Dhall, M.~T. Khan, and R.~A. Shaik, ``Low-complexity
  distributed-arithmetic-based pipelined architecture for an lstm network,''
  \emph{IEEE Transactions on Very Large Scale Integration (VLSI) Systems},
  vol.~28, no.~2, pp. 329--338, 2019.

\bibitem{khan2022architectural}
M.~T. Khan, H.~E. Yant{\i}r, K.~N. Salama, and A.~M. Eltawil, ``Architectural
  trade-off analysis for accelerating lstm network using radix-r obc scheme,''
  \emph{IEEE Transactions on Circuits and Systems I: Regular Papers}, vol.~70,
  no.~1, pp. 266--279, 2022.

\bibitem{alhartomi2023low}
M.~A. Alhartomi, M.~T. Khan, S.~Alzahrani, A.~Alzahmi, R.~A. Shaik,
  J.~Hazarika, R.~Alsulami, A.~Alotaibi, and M.~Al-Harthis, ``Low-area and
  low-power vlsi architectures for long short-term memory networks,''
  \emph{IEEE Journal on Emerging and Selected Topics in Circuits and Systems},
  2023.

\bibitem{pereira2011new}
R.~R. Pereira, C.~H. Da~Silva, L.~E.~B. Da~Silva, G.~Lambert-Torres, and J.~O.
  Pinto, ``New strategies for application of adaptive filters in active power
  filters,'' \emph{IEEE Transactions on Industry Applications}, vol.~47, no.~3,
  pp. 1136--1141, 2011.

\bibitem{li2019prognosis}
Q.~Li, M.~J. Zuo, and S.~Y. Liang, ``Prognosis of bearing degeneration using
  adaptive quaternion least mean biquadrate under framework of hypercomplex
  data,'' \emph{IEEE Sensors Journal}, vol.~20, no.~5, pp. 2659--2670, 2019.

\bibitem{chang2023complete}
C.-Y. Chang, C.-N. Chiou, and S.~M. Kuo, ``A complete design of smart pad that
  reduces snoring,'' \emph{IEEE Transactions on Consumer Electronics}, 2023.

\bibitem{akhtar2023developing}
M.~T. Akhtar, ``Developing a new filtered-x recursive least squares adaptive
  algorithm based on a robust objective function for impulsive active noise
  control systems,'' \emph{Applied Sciences}, vol.~13, no.~4, p. 2715, 2023.

\bibitem{khan2020high}
M.~T. Khan and R.~A. Shaik, ``High-performance hardware design of block lms
  adaptive noise canceller for in-ear headphones,'' \emph{IEEE Consumer
  Electronics Magazine}, vol.~9, no.~3, pp. 105--113, 2020.

\bibitem{khan2017energy}
{Khan, Mohd Tasleem and Shaik, Rafi Ahamed}, ``An energy efficient vlsi
  architecture of decision feedback equalizer for 5g communication system,''
  \emph{IEEE Journal on Emerging and Selected Topics in Circuits and Systems},
  vol.~7, no.~4, pp. 569--581, 2017.

\bibitem{khan2016low}
M.~T. Khan and S.~R. Ahamed, ``Low cost implementation of concurrent decision
  feedback equalizer using distributed arithmetic,'' in \emph{2016 1st India
  International Conference on Information Processing (IICIP)}.\hskip 1em plus
  0.5em minus 0.4em\relax IEEE, 2016, pp. 1--5.

\bibitem{khan2016efficient}
M.~T. Khan, S.~R. Ahamed, and A.~Chatterjee, ``Efficient implementation of
  concurrent lookahead decision feedback equalizer using offset binary
  coding,'' in \emph{2016 20th International Symposium on VLSI Design and Test
  (VDAT)}.\hskip 1em plus 0.5em minus 0.4em\relax IEEE, 2016, pp. 1--6.

\bibitem{allred2003implementation}
D.~Allred, V.~Krishnan, W.~Huang, and D.~Anderson, ``Implementation of an lms
  adaptive filter on an fpga employing multiplexed multiplier architecture,''
  in \emph{The Thrity-Seventh Asilomar Conference on Signals, Systems \&
  Computers, 2003}, vol.~1.\hskip 1em plus 0.5em minus 0.4em\relax IEEE, 2003,
  pp. 918--921.

\bibitem{allred2005lms}
D.~J. Allred, H.~Yoo, V.~Krishnan, W.~Huang, and D.~V. Anderson, ``Lms adaptive
  filters using distributed arithmetic for high throughput,'' \emph{IEEE
  Transactions on Circuits and Systems I: Regular Papers}, vol.~52, no.~7, pp.
  1327--1337, 2005.

\bibitem{guo2011two}
R.~Guo and L.~S. DeBrunner, ``Two high-performance adaptive filter
  implementation schemes using distributed arithmetic,'' \emph{IEEE
  Transactions on Circuits and Systems II: Express Briefs}, vol.~58, no.~9, pp.
  600--604, 2011.

\bibitem{prakash2013low}
M.~S. Prakash and R.~A. Shaik, ``Low-area and high-throughput architecture for
  an adaptive filter using distributed arithmetic,'' \emph{IEEE Transactions on
  Circuits and Systems II: Express Briefs}, vol.~60, no.~11, pp. 781--785,
  2013.

\bibitem{park2013low}
S.~Y. Park and P.~K. Meher, ``Low-power, high-throughput, and low-area adaptive
  fir filter based on distributed arithmetic,'' \emph{IEEE Transactions on
  Circuits and Systems II: Express Briefs}, vol.~60, no.~6, pp. 346--350, 2013.

\bibitem{meher2011high}
P.~K. Meher and S.~Y. Park, ``High-throughput pipelined realization of adaptive
  fir filter based on distributed arithmetic,'' in \emph{2011 IEEE/IFIP 19th
  International Conference on VLSI and System-on-Chip}.\hskip 1em plus 0.5em
  minus 0.4em\relax IEEE, 2011, pp. 428--433.

\bibitem{khan2018optimal}
M.~T. Khan and R.~A. Shaik, ``Optimal complexity architectures for pipelined
  distributed arithmetic-based lms adaptive filter,'' \emph{IEEE Transactions
  on Circuits and Systems I: Regular Papers}, vol.~66, no.~2, pp. 630--642,
  2018.

\bibitem{khan2018improved}
M.~T. Khan, R.~A. Shaik, and S.~P. Matcha, ``Improved convergent distributed
  arithmetic based low complexity pipelined least-mean-square filter,''
  \emph{IET Circuits, Devices \& Systems}, vol.~12, no.~6, pp. 792--801, 2018.

\bibitem{khan2022high}
M.~T. Khan and R.~A. Shaik, ``High-performance vlsi architecture of dlms
  adaptive filter for fast-convergence and low-mse,'' \emph{IEEE Transactions
  on Circuits and Systems II: Express Briefs}, vol.~69, no.~4, pp. 2106--2110,
  2022.

\bibitem{sarma2019novel}
R.~K. Sarma, M.~T. Khan, R.~A. Shaik, and J.~Hazarika, ``A novel time-shared
  and lut-less pipelined architecture for lms adaptive filter,'' \emph{IEEE
  Transactions on Very Large Scale Integration (VLSI) Systems}, vol.~28, no.~1,
  pp. 188--197, 2019.

\bibitem{khan2017low}
M.~T. Khan, S.~R. Ahamed, and F.~Brewer, ``Low complexity and critical path
  based vlsi architecture for lms adaptive filter using distributed
  arithmetic,'' in \emph{2017 30th International Conference on VLSI Design and
  2017 16th International Conference on Embedded Systems (VLSID)}.\hskip 1em
  plus 0.5em minus 0.4em\relax IEEE, 2017, pp. 127--132.

\bibitem{huang2006conjugate}
W.~Huang, V.~Krishnan, and D.~V. Anderson, ``Conjugate distributed arithmetic
  adaptive fir filters and their hardware implementation,'' in \emph{2006 49th
  IEEE International Midwest Symposium on Circuits and Systems}, vol.~2.\hskip
  1em plus 0.5em minus 0.4em\relax IEEE, 2006, pp. 295--299.

\bibitem{jones1993efficient}
D.~L. Jones, ``Efficient computation of time-varying and adaptive filters,''
  \emph{IEEE transactions on signal processing}, vol.~41, no.~3, pp.
  1077--1086, 1993.

\bibitem{huang2011modified}
W.~Huang and D.~V. Anderson, ``Modified sliding-block distributed arithmetic
  with offset binary coding for adaptive filters,'' \emph{Journal of Signal
  Processing Systems}, vol.~63, pp. 153--163, 2011.

\bibitem{wei1986multimemory}
C.-H. Wei and J.-J. Lou, ``Multimemory block structure for implementing a
  digital adaptive filter using distributed arithmetic,'' in \emph{IEE
  Proceedings G (Electronic Circuits and Systems)}, vol. 133, no.~1.\hskip 1em
  plus 0.5em minus 0.4em\relax IET, 1986, pp. 19--26.

\bibitem{chiu1987realization}
Y.-Y. Chiu and C.-H. Wei, ``On the realization of multimemory block structure
  digital adaptive filter using distributed arithmetic,'' \emph{Journal of the
  Chinese Institute of Engineers}, vol.~10, no.~1, pp. 115--122, 1987.

\bibitem{khan2017new}
M.~T. Khan and S.~R. Ahamed, ``A new high performance vlsi architecture for lms
  adaptive filter using distributed arithmetic,'' in \emph{2017 IEEE Computer
  Society Annual Symposium on VLSI (ISVLSI)}.\hskip 1em plus 0.5em minus
  0.4em\relax IEEE, 2017, pp. 219--224.

\bibitem{khan2018area}
{M. T. Khan} and S.~R. Ahamed, ``Area and power efficient vlsi architecture of
  distributed arithmetic based lms adaptive filter,'' in \emph{2018 31st
  International Conference on VLSI Design and 2018 17th International
  Conference on Embedded Systems (VLSID)}.\hskip 1em plus 0.5em minus
  0.4em\relax IEEE, 2018, pp. 283--288.

\bibitem{khan2022two}
M.~T. Khan, M.~A. Alhartomi, S.~Alzahrani, R.~A. Shaik, and R.~Alsulami, ``Two
  distributed arithmetic based high throughput architectures of non-pipelined
  lms adaptive filters,'' \emph{IEEE Access}, vol.~10, pp. 76\,693--76\,706,
  2022.

\bibitem{van2001efficient}
L.-D. Van and W.-S. Feng, ``An efficient systolic architecture for the dlms
  adaptive filter and its applications,'' \emph{IEEE Transactions on Circuits
  and Systems II: Analog and Digital Signal Processing}, vol.~48, no.~4, pp.
  359--366, 2001.

\bibitem{farhang2000analysis}
B.~Farhang-Boroujeny and K.~S. Chan, ``Analysis of the frequency-domain block
  lms algorithm,'' \emph{IEEE Transactions on Signal Processing}, vol.~48,
  no.~8, pp. 2332--2342, 2000.

\bibitem{hazarika2023efficient}
J.~Hazarika, M.~T. Khan, S.~R. Ahamed, and H.~B. Nemade, ``An efficient
  implementation approach to fft processor for spectral analysis,'' \emph{IEEE
  Transactions on Instrumentation and Measurement}, 2023.

\bibitem{hazarika2020area}
{Hazarika, Jinti and Khan, Mohd Tasleem and Ahamed, Shaik Rafi and Nemade,
  Harshal B}, ``An area and power-efficient serial commutator fft with
  recursive lut multiplier,'' in \emph{Modelling, Simulation and Intelligent
  Computing: Proceedings of MoSICom 2020}.\hskip 1em plus 0.5em minus
  0.4em\relax Springer, 2020, pp. 92--100.

\bibitem{hazarika2019energy}
{M.T. Khan, J. Hazarika, S.R. Ahamed, and H.B. Nemade}, ``Energy efficient vlsi
  architecture of real-valued serial pipelined fft,'' \emph{IET Computers \&
  Digital Techniques}, vol.~13, no.~6, pp. 461--469, 2019.

\bibitem{hazarika2019high}
J.~Hazarika, M.~T. Khan, S.~R. Ahamed, and H.~B. Nemade, ``High performance
  multiplierless serial pipelined vlsi architecture for real-valued fft,'' in
  \emph{2019 National Conference on Communications (NCC)}.\hskip 1em plus 0.5em
  minus 0.4em\relax IEEE, 2019, pp. 1--6.

\bibitem{hazarika2019low}
J.~Hazarika, M.~T. Khan, and S.~R. Ahamed, ``Low-complexity continuous-flow
  memory-based fft architectures for real-valued signals,'' in \emph{2019 32nd
  International Conference on VLSI Design and 2019 18th International
  Conference on Embedded Systems (VLSID)}.\hskip 1em plus 0.5em minus
  0.4em\relax IEEE, 2019, pp. 46--51.

\bibitem{srivastava2021fast}
S.~Srivastava, P.~Sharma, S.~Dwivedi, A.~K. Jagannatham, and L.~Hanzo, ``Fast
  block lms based estimation of angularly sparse channels for single-carrier
  wideband millimeter wave hybrid mimo systems,'' \emph{IEEE Transactions on
  Vehicular Technology}, vol.~70, no.~1, pp. 666--681, 2021.

\bibitem{khan2023analyzing}
M.~T. Khan and O.~Gustafsson, ``Analyzing step-size approximation for
  fixed-point implementation of lms and blms algorithms,'' in \emph{2023 IEEE
  Nordic Circuits and Systems Conference (NorCAS)}.\hskip 1em plus 0.5em minus
  0.4em\relax IEEE, 2023, pp. 1--5.

\bibitem{mohanty2012high}
B.~K. Mohanty and P.~K. Meher, ``A high-performance energy-efficient
  architecture for fir adaptive filter based on new distributed arithmetic
  formulation of block lms algorithm,'' \emph{IEEE transactions on signal
  processing}, vol.~61, no.~4, pp. 921--932, 2012.

\bibitem{mohanty2015lut}
B.~K. Mohanty, P.~K. Meher, and S.~K. Patel, ``Lut optimization for distributed
  arithmetic-based block least mean square adaptive filter,'' \emph{IEEE
  Transactions on Very Large Scale Integration (VLSI) Systems}, vol.~24, no.~5,
  pp. 1926--1935, 2015.

\bibitem{khan2018analysis}
M.~T. Khan and R.~A. Shaik, ``Analysis and implementation of block least mean
  square adaptive filter using offset binary coding,'' in \emph{2018 IEEE
  International Symposium on Circuits and Systems (ISCAS)}.\hskip 1em plus
  0.5em minus 0.4em\relax IEEE, 2018, pp. 1--5.

\bibitem{khan2020partial}
M.~T. Khan, J.~Kumar, S.~R. Ahamed, and J.~Faridi, ``Partial-lut designs for
  low-complexity realization of da-based blms adaptive filter,'' \emph{IEEE
  Transactions on Circuits and Systems II: Express Briefs}, vol.~68, no.~4, pp.
  1188--1192, 2020.

\bibitem{khan2021efficient}
M.~T. Khan, R.~A. Shaik, and M.~A. Alhartomi, ``An efficient scheme for
  acoustic echo canceller implementation using offset binary coding,''
  \emph{IEEE Transactions on Instrumentation and Measurement}, vol.~71, pp.
  1--14, 2021.

\bibitem{wallace1964suggestion}
C.~S. Wallace, ``A suggestion for a fast multiplier,'' \emph{IEEE Transactions
  on electronic Computers}, no.~1, pp. 14--17, 1964.

\bibitem{long1989lms}
G.~Long, F.~Ling, and J.~G. Proakis, ``The lms algorithm with delayed
  coefficient adaptation,'' \emph{IEEE Transactions on Acoustics, Speech, and
  Signal Processing}, vol.~37, no.~9, pp. 1397--1405, 1989.

\bibitem{khan2017vlsi}
M.~T. Khan and S.~R. Ahamed, ``Vlsi realization of low complexity pipelined lms
  filter using distributed arithmetic,'' in \emph{TENCON 2017-2017 IEEE Region
  10 Conference}.\hskip 1em plus 0.5em minus 0.4em\relax IEEE, 2017, pp.
  433--438.

\bibitem{arenas2006mean}
J.~Arenas-Garcia, A.~R. Figueiras-Vidal, and A.~H. Sayed, ``Mean-square
  performance of a convex combination of two adaptive filters,'' \emph{IEEE
  transactions on signal processing}, vol.~54, no.~3, pp. 1078--1090, 2006.

\bibitem{khan2018enhanced}
M.~T. Khan and S.~R. Ahamed, ``Enhanced convergence distributed arithmetic
  based lms adaptive filter using convex combination,'' in \emph{2018 Twenty
  Fourth National Conference on Communications (NCC)}.\hskip 1em plus 0.5em
  minus 0.4em\relax IEEE, 2018, pp. 1--6.

\bibitem{baghel2011fpga}
S.~Baghel and R.~Shaik, ``Fpga implementation of fast block lms adaptive filter
  using distributed arithmetic for high throughput,'' in \emph{2011
  International Conference on Communications and Signal Processing}.\hskip 1em
  plus 0.5em minus 0.4em\relax IEEE, 2011, pp. 443--447.

\bibitem{khan2023stochastic}
M.~Khan, O.~Gustafsson \emph{et~al.}, ``Stochastic analysis of lms algorithm
  with delayed block coefficient adaptation,'' \emph{arXiv preprint
  arXiv:2306.00147}, 2023.

\end{thebibliography}
\end{document}